\documentclass[twoside,onecolumn]{article}

\usepackage{mathrsfs,mathtools}
\usepackage[sc]{mathpazo} 
\usepackage[T1]{fontenc} 
\usepackage{microtype} 

\usepackage[hmarginratio=1:1,top=32mm,columnsep=20pt]{geometry} 
\usepackage[hang, small,labelfont=bf,up,textfont=it,up]{caption} 
\usepackage{booktabs} 

\usepackage{lettrine} 

\usepackage{enumitem} 
\setlist[itemize]{noitemsep} 

\usepackage{abstract} 

\usepackage{titlesec} 
\renewcommand\thesection{\Roman{section}} 
\renewcommand\thesubsection{\roman{subsection}} 
\titleformat{\section}[block]{\large\scshape\centering}{\thesection.}{1em}{} 
\titleformat{\subsection}[block]{\large}{\thesubsection.}{1em}{} 

\usepackage{fancyhdr} 
\usepackage{titling} 

\usepackage{hyperref} 

\include{myQcircuit}

\def\transf#1{\mathscr{#1}}

\def\tA{\transf A}
\def\tD{\transf D}

\def\tI{\transf I}
\def\tS{\transf S}
\def\tU{\transf U}
\def\tW{\transf W}

\def\numset#1{\mathbb{#1}}
\def\Nats{\numset{N}}
\def\Int{\numset{Z}}

\def\Reals{\numset{R}}
\def\Cmp{\numset{C}}

\def\bvec#1{\boldsymbol{#1}}
\def\vh{{\bvec h}}
\def\vk{{\bvec k}}
\def\vn{{\bvec n}}
\def\vq{{\bvec q}}
\def\vu{{\bvec u}}
\def\vv{{\bvec v}}
\def\vx{{\bvec x}}
\def\vy{{\bvec y}}
\def\vB{{\bvec B}}
\def\vF{{\bvec F}}
\def\vE{{\bvec E}}
\def\valp{{\bvec \alpha}}
\def\vgam{{\bvec \gamma}}
\def\vsig{{\bvec \sigma}}
\def\vnab{{\bvec \nabla}}

\def\spc#1{\mathscr{#1}}
\def\sH{\spc H}

\def\Spec{\operatorname{Spec}}

\def\<{\langle}
\def\>{\rangle}
\def\kt#1{|#1\rangle}
\def\br#1{\langle#1|}

\pretitle{\begin{center}\Huge\bfseries} 
\posttitle{\end{center}} 
\title {Quantum field theory from first principles} 
\author{%
\textsc{Paolo Perinotti} \\[1ex] 
\normalsize Dipartimento di Fisica, Universit\`a degli Studi di Pavia \\[1ex] 
\normalsize \href{mailto:paolo.perinotti@unipv.it}{paolo.perinotti@unipv.it} 
}
\date{\today} 
\renewcommand{\maketitlehookd}{%
\begin{abstract}
\noindent  
La descrizione matematica dei sistemi quantistici ne identifica univocamente la natura. In 
altre parole, trattiamo un sistema come quantistico se, per descriverne il 
comportamento, adottiamo il formalismo degli spazi di Hilbert e delle relative strutture
individuate dai postulati della teoria quantistica. La scelta di utilizzare sistemi 
quantistici quali sistemi elementari della fisica pu\`o essere giustificata in termini di 
principi informatici, grazie a risultati di ricerche svolte nell'ultimo decennio. Tali 
risultati concludono un programma di ricerca durato quasi un secolo, volto a trovare una 
formulazione della teoria quantistica in termini di principi operazionali. Questo risultato 
pone ora una nuova sfida, che viene qui descritta ed affrontata. Se i sistemi quantistici 
sono in prima battuta pensati come portatori elementari di informazione, piuttosto che come 
costituenti elementari della materia, e le loro connessioni sono connessioni logiche 
all'interno di un algoritmo, piuttosto che relazioni spazio-temporali, occorre trovare una 
opportuna giustificazione per i concetti meccanici--- che caratterizzano la teoria 
quantistica come teoria dei sistemi fisici. A tal fine, illustreremo come si possa pensare 
una legge fisica quale algoritmo di aggiornamento del contenuto dei registri di memoria che 
compongono un sistema. Imponendo quindi le propriet\`a caratteristiche delle leggi fisiche 
a tale algoritmo, ovvero omogeneit\`a, reversibilit\`a ed isotropia, mostreremo che le 
leggi fisiche cos\`\i\ selezionate sono particolari algoritmi noti sotto il nome di automi 
cellulari. Ulteriori assunzioni di massima semplicit\`a dell'algoritmo portano a due soli 
automi cellulari, che in un opportuno regime possono essere descritti dalle equazioni 
differenziali di Weyl, alla base della dinamica di campi quantistici relativistici. 
Discuteremo infine come lo stesso automa cellulare possa dar luogo tanto alla dinamica di 
particelle Fermioniche, quanto alle leggi di Maxwell, che regolano la dinamica del campo 
elettromagnetico. Concluderemo con la discussione del principio di relativit\`a, che, 
opportunamente riformulato, consiste nella definizione del concetto di cambio di sistema di 
riferimento inerziale, e che, opportunamente rielaborato, permette di ritrovare la 
simmetria dello spazio-tempo Minkowskiano, definita dal gruppo di Poincar\'e. Lo spazio-
tempo emerge quindi in tale contesto, non come sfondo per le leggi fisiche, ma come loro 
manifestazione, con caratteristiche dettate dalla dinamica dei sistemi, inevitabilmente 
rivestito delle equazioni che descrivono tale dinamica. In breve, non esiste spazio-tempo 
senza un'equazione di evoluzione che lo richieda.
\\
\\
The mathematical description of quantum  systems univocally identifies their nature. 
In other words we treat a system as quantum if we describe its behaviour adopting 
Hilbert spaces and structures thereof, as prescribed by the postulates of quantum theory.
The choice of using quantum systems as the elementary systems of physics can be justified 
in terms of informational principles, thanks to results of the last decade. Such
results come as the conclusion of a research program that lasted almost one century, 
with the aim of reformulating quantum theory in terms of operational principles.
This achievement now poses a new challenge, that we face here. If the systems of 
quantum theory are thought of as elementary information carriers in the first place, 
rather than elementary constituents of matter, and their connections are logical 
connections within a given algorithm, rather than space-time relations, then we need to
find the origin of mechanical concepts---that characterise quantum mechanics as a 
theory of physical systems. To this end, we will illustrate how physical laws can be 
viewed as algorithms for the update of memory registers that make a physical system.
Imposing the characteristic properties of physical laws to such an algorithm, 
i.e.~homogeneity, reversibility and isotropy, we will show that the physical laws thus 
selected are particular algorithms known as cellular automata. Further assumptions 
regarding maximal simplicity of the algorithm lead to two cellular automata only, that in
a suitable regime can be described by Weyl's differential equations, lying at the basis
of the dynamics of relativistic quantum fields. We will finally discuss how the same 
cellular automaton can give rise to both Fermionic field dynamics and to Maxwell's
equations, that rule the dynamics of the electromagnetic field. We will conclude reviewing 
the discussion of the relativity principle, that must be suitably adapted to the scenario
where space-time is not an elementary notion, through the definition of a change of
inertial reference frame, and whose formulation leads to the recovery of the symmetry of 
Minkowski space-time, identified with Poincar\'e's group. Space-time thus emerges as 
one of the manifestations of physical laws, rather than the background where they occur, 
and its features are determined by the dynamics of systems, necessarily equipped with 
differential equations that express it. In brief, there is no space-time unless an
evolution rule requires it.
\end{abstract}
}

\begin{document}

\maketitle
	The advent of Quantum Mechanics in the early twentieth century probably represents the
most profound revolution in the history of natural sciences, surely comparable to 
the Copernican revolution or to the introduction of the Galilean method, and in most respects even more shaking than the development of general relativity. The consequences 
of the quantum world-view are still far from being grasped in their full extent, even
by the community of physicists themselves. This is true if one accepts the Copenhagen 
interpretation of quantum physics, as well as if one denies it and yearns for a consistent
way of reconciling quantum theory with a classical imaginative world.

While there are attempts at formulations of quantum mechanics that can save a notion of predetermined elements of reality independent of observations, we will focus here on the research program that comes from acceptance of a world-view where physical events occur in an inherently probabilistic way, depending on the choices of observers that measure them, and the best physical theory we can wish for must provide rules for calculating correlations between possible events. In this scenario, concepts like mass, charge, or temperature of a system, are high-level structures as compared to registrations of detector clicks or pointer positions, or naked-eye observations. A further thought on the situation then leads to the conclusion that our bottom-level theory is rather a theory of information than a mechanical theory: information about operations performed on systems and about observed events. Mechanics is the high-level picture that we wish to recover at the end of our route towards a reformulation of fundamental physics.

It is then clear why most concepts, techniques and tools that we use in this endeavour are borrowed from that special experience in the history of quantum science that is quantum information theory. Since its very beginning, besides introducing new information-processing concepts and technologies, quantum information theory has represented a new way of looking at foundations of Quantum Theory (QT). The reconsideration of the structure of quantum theory and the exploration of areas that were neglected before, lead to a new axiomatization program, initiated in the early 2000 
\cite{hardy2001quantum,Fuchs:2002p89,dariano114,d2010probabilistic} and lasted for more than one decade~\cite{dakic2009quantum,masanes2011derivation,hardy2011reformulating}. The goal of the endeavour was to reconstruct von Neumann's mathematical formulation of the theory in terms of Hilbert-spaces---neglecting the mechanical postulates such as Schr\"odinger's equation or its relativistic counterparts---starting from information-processing principles. A complete derivation of QT for finite dimensional systems has been finally achieved in Ref.~\cite{quit-derivation} within the framework of Operational Probabilistic Theories (OPT), starting from six principles assessing the possibility or impossibility to carry out specific information-processing tasks. A book has later been published~\cite{bookDCP2017} that presents the framework of OPTs---theories that share the same structures for composition of systems and processes as QT and classical theory, and include suitable rules for calculating probabilities of composite processes---along with the derivation of QT.

Coming back to our metaphor, however, at this stage we are not even half-way through the journey from the low-level theory of physical systems to the high-level language of mechanics. The informational approach is then carried one step further, with the purpose of reconstructing quantum equations of motion, and space-time as their manifestation.
In the simplest, non interacting case, we will recover Weyl, Dirac and Maxwell free field theories, along with the fundamental constants they involve, such as $\hbar$ and $c$. 

The key idea from this stage on is to think of physical laws as information processing algorithms, which process the state of an array of quantum memory cells. If one follows the original proposal by Feynman~\cite{feynman1982simulating}, where physical laws are supposed to be amenable to an {\em exact} simulation by an algorithm that requires a bounded amount of resources per unit volume of space-time, then the distinction between simulation algorithm and actual physical law vanishes\footnote{Differently from the proposal of Feynman, we consider here physical laws exclusively as scientific tools to analyse physical phenomena, and distinguish them, in principle, from the actual rules governing the occurrence of phenomena. It is clear, however, that the goal of the formulation of physical laws is to predict the behaviour of physical systems.}. In this case, the information capacity of a finite physical system has to be bounded. From this perspective, then, any model of physical law will correspond to an algorithm where 
continuous quantum fields will be replaced by countably many finite-dimensional Fermionic\footnote{The motivation for the choice of Fermionic systems rather than qubits would require a very long discussion per se. Here we will only remark that while qubit algorithms are easily simulated by Fermionic algorithms, respecting their topological structure, for the converse simulation one needs to distort the topology of connections between elementary processes. Moreover, the requirement of linearity that we will discuss shortly is not satisfied by non-trivial qubit algorithms. For obvious reasons, Bosonic systems would violate the requirement of a bounded information density.} quantum systems. Such an algorithm, reversible and abiding by some weak form of homogeneity and locality, then corresponds to a Fermionic Cellular Automaton.
Cellular automata, originally introduced in classical computation by von Neumann~\cite{VonNeumann66}, were studied in the quantum domain since the late eighties of last century~\cite{grossing1988quantum,aharonov1993quantum,ambainis2001one}. It was only in 2004 that the concept was rigorously formalised~\cite{schumacher2004reversible}, and in the following we will use the term Quantum Cellular Automaton (QCA) to refer to this notion. Now the subject has been studied by various authors in further detail~\cite{Gross2012,arrighi2011unitarity,Freedman:2019wt,Arrighi:2019wq,Farrelly:2019vr}. Very recently, the author formalised cellular automata in the wider context of OPTs~\cite{perinotti2019cellular}.

It is due to mention that studies similar in their technical development have been carried out before. The simulation of relativistic quantum dynamics via QCAs, in particular, was already discussed in Refs.~\cite{bialynicki1994weyl,meyer1996quantum,Yepez:2006p4406}. 
However, we remark that the program that we are discussing here adopts a paradigm that is reversed with respect to those mentioned above, and aims at a derivation of quantum field theory (QFT) from an informational standpoint~\cite{PhysRevA.90.062106,DAriano:2016aa}. Other authors, following these results, also addressed the foundations of QFT in the QCA framework~\cite{Arrighi:2019wq,Farrelly:2019vr}.

The present status of this research program is very successful, having derived from a few, very simple requirements, the equations of Weyl and Dirac in 1+1~\cite{bisio2013dirac,Bisio2015244} and in 3+1 dimensions~\cite{PhysRevA.90.062106}, along with Maxwell's equations~\cite{BISIO2016177} for suitable bilinear functions of the Weyl fields, also abiding by bosonic commutation relations to very good approximation. 
Weyl's, Dirac's and Maxwell's equations are the dynamical equations of the fundamental relativistic quantum fields. The above results then imply that the free, non-interacting fields describing the most elementary physical systems have been recovered.

In the same project we can list a stream of works that analyse the symmetries of the physical laws 
represented by the above mentioned QCAs~\cite{MAURODARIANO2013294}, starting from a rigorous 
application of the relativity 
principle~\cite{refId0,PhysRevA.94.042120,Bisio:2017aa,apadula2018symmetries}. The results of these 
analyses show that the relativistic Minkowski space-time can be recovered, as the natural manifold 
for the representation of physical laws represented by the above mentioned QCAs. Indeed, while the 
discrete nature of QCAs manifestly breaks covariance under the usual representations of Lorentz's 
group, upon identifying the notion of ``reference frame'' with that of ``representation'' of the 
dynamics, one can appeal to the relativity principle to define the ``inertial representation'' as the 
one for which the physical law retains the same mathematical form. In such a way the change of 
inertial reference frame leads to a set of modified Lorentz transformations that recover the usual 
ones when the observation scale is much larger than the discrete microscopic scale.

The future challenge consists in accounting also for the fundamental interactions between elementary fields. The program presents with technical difficulties, and preliminary studies have been carried out, analysing 1+1-dimensional toy-theories~\cite{PhysRevA.97.032132,Bisio_2018}, and introducing perturbative techniques that genralise their counterpart for continuous-time, Hamiltonian dynamics~\cite{aless2019scattering}.

\section{Cellular automata}

The approach illustrated in the introduction is based on the assumption that physical systems and 
events are the manifestation of information processing occurring on an array of elementary memory
cells. This paradigm is that of a general algorithm running on a computer, whose memory registers
are systems of some yet unspecified kind. They might be classical bits---{\em bits} for short
---or larger classical systems, or quantum bits---{\em qubits}---or Fermionic modes, or even 
systems of some more exotic kind. The choice of the type of system that we use for our model
of physical law can be motivated in various ways, but will be justified, in the end of the day,
only {\em  posteriori}, based on its predictions. It is reasonable to suppose, however, that our
theoretical model should not be unnecessarily complex, according to a loose actualisation of
Occam's razor. Following Feynman~\cite{feynman1982simulating} and 
Deutsch~\cite{deutsch1985quantum} we then suppose that the physical laws should be exactly 
replicated by a computer operating on finite means as long as a finite portion of space-time is to 
be described. 
Reasoning in this way, one can imagine that every elementary system should carry
only a finite amount of information. While the last assumption is not strictly necessary, its 
violation seems to be at odds with the existence of fundamental physical constants such as $c$, 
$\hbar$ and $G$, that determine spatial scales separating regimes where physical systems behave
in a very different way. The choice that we make is then for a kind of system with a finite 
number of levels.

Our choice for the elementary cells will be be quantum, because present day physics suggests that 
the behaviour of elementary physical systems is consistent with such a picture. However, we have
to choose between the two simplest, finite-dimensional quantum systems: qubits or Fermionic 
modes. For the moment it is not strictly necessary to exclude one of the two, and we will come
back on this question later.

The big question now is: what is the best candidate algorithm for representing physical laws? More
precisely, what features must it cast? What are the constraints that we impose on our algorithm?
While, also in this case, our assumptions are not strictly necessary, we analyse the features
of a physical law, and use them as desiderata for our algorithm. In the first place, physical 
laws are homogeneous in space-time. The equations that express them hold independently of the
space-time point that we are considering. How can we translate this requirement for our 
algorithm, as we do not allow space-time to enter our basic language? Homogeneity in our
computational scenario is the requirement that, roughly speaking, every memory system is treated
in the same way by the algorithm, and the rule for updating the content of a cell in a 
computational step is then independent of both i) the cell address, and ii) the step 
counter of the algorithm. We formulate the homogeneity principle as follows

\medskip 

{\em Homogeneous update rule}: every two memory cells cannot be distinguished by the way in which 
the rule updates their state, unless one establishes a reference cell, which can be any.

\medskip

In order to make the content of the principle clearer, let us think of a memory cell as a point 
in space (or space-time). The physical laws are such that one cannot set an absolute reference 
frame, where a special point $p$ is chosen as the origin, based on a special behaviour of 
physical laws at $p$. The first part of the principle then extends this property to an abstract
memory array evolved by an update rule. On the other hand, if any reference frame is fixed in 
space (or space-time), then it is possible to label points in space (or space-time) with respect 
to such a reference frame. Einstein's construction of an inertial frame via clock synchronisation 
is a constructive example of the desideratum that physical laws allow an observer to distinguish the 
physical role of points in space-time with respect to a reference point. The possibility of a similar 
construction is bound to the possibility of operationally discriminate two points, so that their
``geometric label'' in space-time, or their logical address in a memory array is not merely 
theoretical. In other words, it is meaningful to name two different cells by different labels, once
we arbitrarily set a reference frame. This requirement is captured and generalised by the second part 
of the principle.

The second property of physical laws that we require for our update rule is reversibility. 
Despite its intuitive content, there might be different meanings for the reversibility 
requirement. The first meaning comes from the framework of OPTs where an agent is supposed
to be able to perform on a system every transformation that is mathematically conceivable and not 
inconsistent. Then, a transformation is reversible if there exists another one that inverts its 
effects, and whose effect is inverted by it. This concept of reversibility, that we name {\em 
operational invertibility}, is clearly not applicable to a physical law, since there is no 
possibility of changing a physical law to its inverse {\em in practice}. On the opposite side, one 
has a notion of reversibility related to the possibility to use knowledge of the state of a cell or a 
group of cells to mathematically reconstruct the state of another group of cells at an earlier 
evolution step. We call this property {\em invertibility}. Invertibility might seem a reasonable 
requirement for a physical law, but we need to remind that in the theories we are considering it 
is in principle impossible to acquire the knowledge of the state of any group of 
cells~\cite{single}. The last notion that we discuss, that we call reversibility, lies in between 
the above two, and is the notion that is commonly adopted in quantum mechanics textbooks. 
According to this notion, an evolution is reversible if it is equivalent to its inverse modulo 
operationally reversible transformations acting on local memory cells. 

Suppose now that we have an invertible evolution rule $\tU$, whose mathematical inverse 
$\tU^{-1}$ is another possible evolution rule. Suppose that the two rules are not reversible in 
the sense defined above. It is very easy to construct a reversible rule by taking two copies of 
the memory array, thus building a ``two-layer'' memory with addresses $(a,i)$, where $a\in\Nats$ 
is the address of a cell in the single layer, while $i\in\{0,1\}$ addresses the layer. The  
cell $a$ in the new memory array is the composite systems made of the cells $(a,0)$ and $(a,1)$.
Let then the evolution rule on this new array be $\tU\otimes\tU^{-1}$. This rule is reversible, 
since it is equivalent to $\tU^{-1}\otimes\tU$ modulo swapping the systems $(a,0)$ and $(a,1)$ 
for every $a$. The latter operation is a local, operationally reversible transformation on the cell 
$a$. The formulation of reversibility is very simple.

\medskip

{\em Reversible update rule}: the update rule is reversible, i.e.~it is equivalent to its 
inverse via one and the same operationally reversible transformation acting independently on 
every cell.

\medskip
  
Let us then consider, in what follows, a denumerable memory array, with a reversible update rule 
that is homogeneous. We make a final requirement on our update rule, that is {\em locality}.
If we started considering systems in a geometrically defined space-time, we could define locality
as the property of an update rule to change the state of a system in a way that only depends on 
``close'' systems, where closeness is defined by the underlying geometry. However, in our 
approach we are not introducing physical concepts at the fundamental level, and we are trying
to show that the latter can be recovered as emergent features, including the geometry
of space-time. Thus, we have to turn the usual approach to locality on its head, and define
two memory cells $a$ and $b$ to be {\em neighbours} if the update rule is such that either a 
change in $a$ at step $t$ affects $b$ at $t+1$, or viceversa. Thus, the distance of two cells 
$a,b$ is defined by the number $d$ of steps that are required for a change in $a$ (or $b$) at step 
$t$ to affect $b$ (or $a$) at $t+d$. 

\begin{figure}[h]
\centering
\includegraphics[width=12cm]{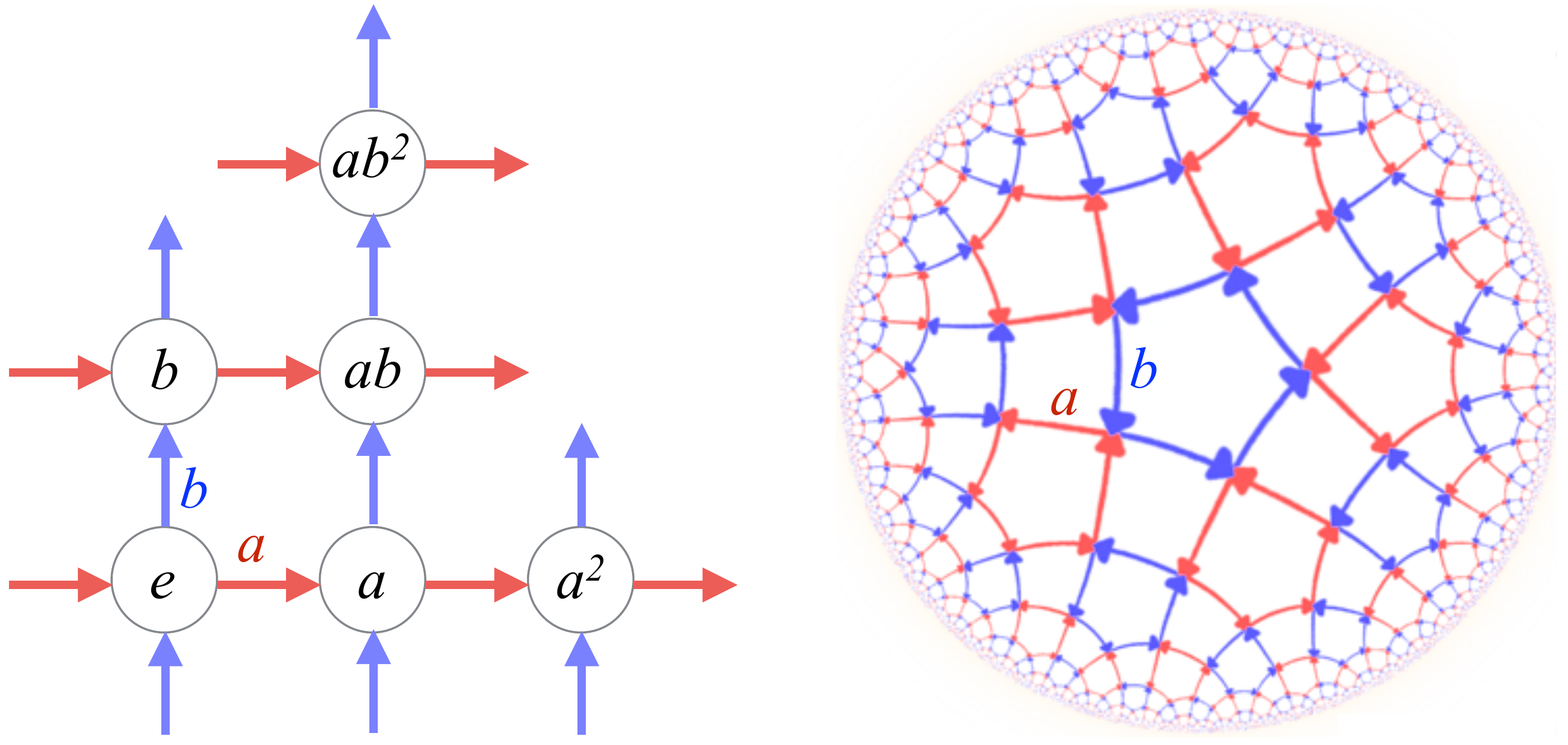}
\caption{Two examples of a Cayley graphs. On the left: a Cayley graph $\Int^2$ with two 
generators $a,b$. The only non-trivial relation is the Abelianity condition $aba^{-1}b^{-1}$.
On the right: the group is a Fuchsian group presented as $\<a,b\mid a^5,b^5,abab\>$. The graph is 
quasi-isometrically embedded in the Poincar\'e disc, which makes it quasi-isometric to the 
hyperbolic plane $\numset H_2$.}
\label{f:cay}
\end{figure}

The geometry of our memory array can be made more explicit by constructing a graph $(V,E)$, as 
follows. The vertex set is $V\equiv A$, the set of all possible addresses in our array, while the 
(ordered) edge set $E$ contains all pairs $(a,b)\in A\times A$ such that a change in $a$ at step 
$t$ can affect $b$ at $t+1$. For a homogenous update rule, the graph $(V,E)$ is very regular: 
being all the cells identical, their neighbourhoods must have the same structure: one can then 
choose a reference cell $e$, and identify its neighbour cells by the symbols 
$\{h_i\}_{i\in S_+}$. The same set then identifies the neighbours of every other cell $a$. One 
can prove, moreover, that the graph of a homogeneous rule is the Cayley graph of a 
group~\cite{DAriano:2016aa,perinotti2019cellular}. A Cayley graph of a group $G$, presented by a set 
of generators $S_+=\{h_i\}_{i\in S_+}$ along with a sufficient set of relations\footnote{A set of 
relators is a set of group elements of the form $h_{i_1}^{s_1}h_{i_2}^{s_2}\ldots h_{i_k}^{s_k}$, 
with $s_k\in\{+1,-1\}$, that are set equal to the group unit $e$. The set $R$ is sufficient if every 
product of generators that is equal to $e$ belongs to the conjugate closure of $R$.} $R$, is a graph 
denoted as $\Gamma(G,S_+)$, whose vertex set is $G$, and the edge set $E\subseteq G\times G$ is 
defined as $E=\{(g,g')\mid g\in G,\ g'=gh_i,\ i\in S_+\}$. Notice that the edges of a Cayely graph 
are colored, the color of an edge $(g,gh_i)$ corresponding to the generator $h_i\in S_+$ (see 
fig.~\ref{f:cay}). Moreover, the Cayley graph is identified by the neighbourhood structure and by 
any sufficient set of closed paths representing the elements of $R$.

We now want to use the graph representing causal relations of an update rule to define rigorously
the notion of locality. In order to make this principle independent of homogeneity, we consider
a graph that has no further structure, in particular it needs not be a Cayley graph.
The notion of distance that we defined above coincides with the definition of distance on the 
graph. Let $p\coloneqq(a_1,a_2,\ldots,a_n)$ be a path in $A$ i.e. a collection of vertices such 
that either $(a_i,a_{i+1})\in E$ or $(a_{i+1},a_i)\in E$, and $a_1=a$, $a_n=b$. The set of all 
these paths is $\gamma(a,b)$. Then the distance of $a$ and $b$ on the graph will be defined as

\begin{align}
d(a,b)\coloneqq\min_{\gamma(a,b)}l(p).
\end{align}

With the above definition of distance, it is tautological that the update rule is local, provided 
we adopt a naive notion of locality, i.e.~the evolution is such that changes on a cell affect only 
neighbouring cells in a single step. However, we want our notion of locality to be such that 
exceedingly connected cells are forbidden: the neighbourhood of a cell $a$ must be  of bounded 
size. More than that, since the memory array is allowed to be infinite, we want the bound to be 
uniform. This avoids the possibility that every single cell has a finite number of neighbours, 
but overall the neighbourhood size is unbounded. We observe that the last remark is important 
only because we want to define locality independently of homogeneity. Before concluding the 
discussion of locality, an important remark is in order. Up to now we only referred to pointwise 
properties of the update rule, i.e.~properties ruling the behaviour at a given cell or in a small 
neighbourhood of cells. This is in line with the usual notion of a physical law, that is 
typically formulated as some differential equation holding at each and every single point in 
space-time, and, as such, it can be tested with local experiments. In other words, local 
experiments are sufficient to gather enough information about physical laws. In our framework, in 
order for the latter statement to hold true, locality and homogeneity are not sufficient. Indeed, 
in order to specify an update rule completely, the large scale structure of the graph has to be 
specified. In other words, knowledge of the neighbourhood of any point and the rule that updates it 
must be supplemented by the specification of closed paths. The locality requirement then amounts to 
require that every closed path can be decomposed into ``elementary'' closed paths that have a 
uniformly bounded length. For a homogeneous rule, this amounts to say that a sufficient set $R$ of 
relations is finite, and the length of its elements is bounded. The latter condition grants us the 
possibility of testing physical laws on regions that have a bounded size.

\medskip

{\em Local update rule}: the size of the neighbourhood of every cell is uniformly bounded, and
closed paths on the graph of the update rule are decomposable into elementary closed paths of 
uniformly bounded length.

\medskip

Now, a memory array with a homogeneous, reversible and local update rule is a {\em Cellular 
Automaton} (CA). For a more formal definition, the reader is referred to 
Ref.~\cite{perinotti2019cellular}.

We now want to stress that the metric space defined by the Cayley graph $\Gamma(G,S)$ with the 
distance $d$, besides carrying all the algebraic information about $G$, carries geometric 
information. In particular, it identifies an equivalence class of metric spaces that contains all
the Cayley graphs of $G$, as well as other metric spaces including suitable smooth manifolds, where the Cayley graphs of $G$ can be embedded with a uniformly bounded distortion of distances.
The geometric information about $G$ captured by such an equivalence class is generally bound to its 
algebraic properties. Rigorously speaking, the equivalence relation we are referring to is {\em 
quasi-isometry}. Let $(M_1,d_1)$ and $(M_2,d_2)$ be two metric spaces. We say that a map 
$f:M_1\to M_2$ is a quasi-isometry if one can find real constants $1<a<\infty$ and $0<b<\infty$ such 
that for every $x,y\in M_1$
\begin{align}
\frac1ad_2(f(x),f(y))-b\leq d_1(x,y)\leq a d_2(f(x),f(y))+b,
\end{align}
and a real constant $0<k<\infty$ such that for every $z\in M_2$ there exists $x\in M_1$ so that
\begin{align}
d_2(z,f(x))\leq k.
\end{align}
If such a map $f$ exists, we say that $M_2$ is quasi-isometric to $M_1$. Quasi-isometry is easily 
shown to be an equivalence relation. Two quasi-isometric spaces can be very different, e.g.~one can 
be discrete, like a graph, and the other can be smooth, like a differentiable manifold. However, they 
have some common intrinsic geometrical property. Points in two quasi-isometric spaces can be 
associated in a way that ``almost'' preserves  distances. In this sense, every CA naturally 
identifies a class of metric spaces, typically including smooth manifolds, through their Cayley graph 
(see e.g.~fig.~\ref{f:cay}).
This class bears the relevant geometric information about the structure of space (and also space-time) that emerges from the physical law represented by the automaton. The theory that deals with quasi-isometry classes and the connection they allow one to draw between algebraic and geometric properties of groups is called {\em Geometric group theory}. Curiously, this theory seems to fill the gap lamented by A.~Einstein in the following quote, which is amazing, yet not among the most famous of his:

\medskip 

{\em But you have correctly grasped the drawback that the continuum brings. If the molecular view 
of matter is the correct (appropriate) one, i.e., if a part of the universe is to be represented 
by a finite number of moving points, then the continuum of the present theory contains too great 
a manifold of possibilities. I also believe that this too great is responsible for the fact that 
our present means of description miscarry with the quantum theory. The problem seems to me how 
one can formulate statements about a discontinuum without calling upon a continuum (space-time) 
as an aid; the latter should be banned from the theory as a supplementary construction not 
justified by the essence of the problem, which corresponds to nothing ``real''. But we still lack 
the mathematical structure unfortunately. How much have I already plagued myself in this 
way!}~\cite{stachel2001einstein}.

\medskip

\section{Linear Fermionic cellular automata and quantum walks}

A Fermionic CA has a memory array that is made of local Fermionic modes, whose
transformations, states and effects can be expressed in terms of the Fermionic algebra generated 
by the local field operators $\psi_s(a)$ with their adjoints $\psi^\dag_s(a)$, where $a\in A$ and 
$s\in H_a$ is an index that accounts for the internal structure of the cell $a$, which is 
composed by $|S_a|$ local Fermionic modes. The algebra is fully characterised by the Canonical 
Anticommutation Relations (CAR)
\begin{align}
\{\psi_s(a),\psi^\dag_t(b)\}=\delta_{ab}\delta_{st}I,\quad\{\psi_s(a),\psi_t(b)\}=0,
\end{align}
where $\{X,Y\}\coloneqq XY+YX$ denotes the {\em anticommutator} of $X$ and $Y$.

For homogeneous rules, the cell structure $H_a=H$ is independent of $a$. A reversible map $\tA$ 
of a Fermionic memory array is defined as an automorphism of the algebra of operators, and it is 
then fully specified by the image of field operators $\tA[\psi_s(a)]$. In the general case, the 
latter is a polynomial in the field operators of the neighbourhood of the cell $a$. Normally, we 
would require an update rule that cannot create excitations, that is to say,  a number-preserving 
automorphism:
\begin{align}
\tA(N)=N,
\end{align}
where $N\coloneqq\sum_{a,s}\psi^\dag_s(a)\psi_s(a)$. In the following we will make the further
restrictive assumption that the automorphism is {\em linear}, i.e.~
\begin{align}
\tA[\psi_s(a)]=\sum_{h_i\in S}W_{ss'}(h_i)\psi_{s'}(ah_i^{-1}),
\end{align}
where $S\coloneqq S_+\cup S_-$, and $S_-\coloneqq S^{-1}$. The matrix $W(h_i)$ with elements 
$W_{ss'}(h_i)$ is called the {\em transition matrix} corresponding to the generator $h_i\in S$. 
Since the automaton is linear, the transition matrix contains all the information about the 
dynamics of the system. In Fermionic theory, a basis of pure states, identified by the string 
$\bvec a_{\bvec s}=(a_1s_1,a_2s_2,\ldots ,a_ks_k)$ can be defined by
\begin{align}
\kt{\bvec a_{\bvec s}}\coloneqq\psi^\dag_{s_1}(a_1)\psi^\dag_{s_2}(a_2)\ldots\psi^\dag_{s_k}(a_k)\kt\Omega,
\end{align}
where $\kt\Omega$ is the unique common eigenvector of the (commuting) number operators 
$\psi^\dag_{s}(a)\psi_{s}(a)$ with null eigenvalues for every $a$ and $s$. By some 
straightforward algebra, one can easily realise that the transition matrices $W^*(h_i)$
define the evolution of states of a single excitation
\begin{align}
\kt{\psi(t)}=\sum_{a\in A,s\in H}\varphi_s(a,t)\kt{as},
\end{align}
namely
\begin{align}
\kt{\psi({t+1})}=\sum_{a\in A,s,s'\in H,h_i\in S}W^*_{ss'}(h_i)\varphi_s(a,t)\kt{ah_i^{-1}s'},
\end{align}
in other words
\begin{align}
\varphi_s(a,t+1)=\sum_{s'\in H,h_i\in S}W^*_{ss'}(h_i)\varphi_s(ah_i^{-1},t).
\end{align}
The dynamics of single excitations thus determines the full dynamics, and multiple excitations 
evolve independently. Treating $\varphi_s(a,t)$ as a ``single particle'' wavefunction in 
$\sH=l^2(A)\otimes \mathbb C^{|H|}$, the space of square summable sequences on $A\times H$, for 
groups having suitable finite quotients the transition matrices define a unitary 
operator\footnote{In the general case, preservation of the Fermionic algebra easily allows one to 
prove that $W^\dag$ is an isometry, however the proof of unitarity can be carried out for CAs on 
Cayley graphs of groups satisfying the {\em wrapping lemma} (see 
Refs.~\cite{schumacher2004reversible,perinotti2019cellular}).} $W$ on 
$\sH$~(for details see Ref.~\cite{PhysRevA.98.052337}). One can easily 
verify~\cite{PhysRevA.90.062106,PhysRevA.100.012105} that $W$ can be expressed as
\begin{align}
W=\sum_{h_i\in S}T_{h_i}\otimes W(h_i),
\label{eq:walk}
\end{align}
where $T_{g}$ is the {\em right-regular} representation of the group $A$, i.e.~
$T_g\kt f\coloneqq\kt{fg^{-1}}$, with $\{\kt f\mid f\in A\}$ the canonical orthonormal basis 
in $l^2(A)$ and $g\in A$. A unitary operator of the form~\ref{eq:walk} on 
$l^2(A)\otimes\mathbb C^{|H|}$ is called {\em Quantum Walk} (QW), because it represents the 
coherent quantum version of a classical random walk.

We remark that not only a linear Fermionic CA can be represented by a QW, but also
every QW corresponds to a linear Fermionic CA. Indeed, consider a general QW $W$ as in 
Eq.~\eqref{eq:walk}, and define the linear endomorphism of the Fermionic algebra on the 
corresponding Cayley graph
\begin{align*}
\tA[\psi_s(g)]\coloneqq \sum_{h_i\in S}W_{ss'}(h_i)\psi_{s'}(gh_i^{-1}).
\end{align*}
Thanks to unitarity of $W$, the endomorphism $\tA$ is actually an automorphism, and one can easily 
show that it corresponds to a Fermionic CA.

\section{Isotropy}

In this section we discuss a further principle that we impose on our CAs: isotropy.
Once again, while the name of the property is borrowed from space-time geometry, in order to
give it a meaning in a pre-geometric context---where we only have a memory array organised as a
Cayley graph by virtue of causal connections introduced by a dynamical law---we need to rethink
the notion of isotropy in depth, and abstract its essence avoiding to appeal to a background
geometry. Notice that in the case of isotropy avoiding {\em every} geometrical notion is
impossible, but we will strictly refer to those introduced by homogeneity, i.e.~ the Cayley
graph representing causal connections of memory cells and its properties. The notion of isotropy,
in space-time, requires every direction to be equivalent. The translation fo this principle
in our context leads to the requirement that, given the Cayley graph $\Gamma(A,S)$, every
generator in $S_+$ is equivalent. This means that, if we permute the labels of elements in $S_+$
according to some transitive permutation group, we will observe a physical law that is 
equivalent to the original one. Equivalence of the two physical laws is intended, as in the case 
of reversibility, through one and the same operationally reversible transformation acting 
independently on every cell.

Let us now spell out isotropy in rigorous terms. Let  $L$ be a group of automorphisms of the 
Cayley graph $\Gamma(A,S)$ (i.e.~permutations of the vertices that map edges to edges) that can 
be expressed as a permutation $\lambda$ of $S$. This means that, if $l\in L$, for every 
$a=h_{i_1}^{s_1}h_{i_2}^{s_2}\ldots h_{i_k}^{s_k}\in A$ one has 
$l(a)=\lambda(h_{i_1}^{s_1})\lambda(h_{i_2}^{s_2})\ldots\lambda(h_{i_k}^{s_k})$. 
It is easy to verify that a Cayley graph isomorphism that 
can be expressed as a permutation is also a group automorphism of $A$. We say that a quantum walk 
on $\Gamma(A,S)$ is $L$-isotropic if there exists a projective unitary faithful representation 
$U:L\to\mathcal U(\mathbb C^{|H|})$ such that
\begin{align}
W_{l(h_i)}=U_lW_{h_i}U_l^{-1},
\end{align}
for $l\in L$ and $W_{l(h_i)}\neq W_{h_i}$, with the group $L$ acting transitively on $S_+$. 
This notion allows us to formulate a new principle, which we call {\em isotropy} (see 
ref.~\cite{PhysRevA.100.012105}).

\medskip

{\em Isotropic update rule}. A Quantum Walk is isotropic if it is $L$-isotropic for some group $L$
of automorphisms of the Cayley graph $\Gamma(A,S)$ that can be expressed as a permutation 
$\lambda$ of $S$.

\medskip

In the remainder, in addition to the homogeneity, reversibility and locality requirements, we 
will restrict to CAs that are linear and whose QW is isotropic.

\section{Quantum Walks on Cayley graphs of $\Int^d$}

We will now restrict attention to Cayley graphs of $\Int^d$. Up to now our approach was fully
deductive, i.e.~we analysed the simple consequences of our requirements on the update rule for
our memory array. This, in fact, is the spirit of our approach. However, in this section we will
reason in an opposite way. We start from simplest non-trivial cell structure ($|H|=2$) and from 
the group (i.e.~$\Int$), and ask ourselves what of its Cayley graphs can support an isotropic 
linear Fermionic CA. Notice that the simplest cell structure corresponds to $|H|=2$, 
because for Cayley graphs of Abelian groups one can prove~\cite{PhysRevA.93.062334} that for 
$|H|=1$ one has only the trivial QW. Before restricting to the interesting physical case of 
$d=3$, we discuss general properties of QWs on Cayley graphs of Abelian groups that allow us
to simplify the analysis of the unitarity constraints and to make the problem of diagonalisation
much simpler.

Indeed, the right-regular representation of Abelian groups is Abelian, and this implies that
the operators $T_{g}$ can be simultaneously diagonalised in the improper basis $\kt\vk$.
The distributions $\kt\vk$ are easily computed after we introduce the basis $\tilde h_j$ defined
as follows. First, we choose in $\Int^d$ a canonical basis, and express every $h_i\in S$ as a
vector $\vh_i$. We then define the sets $D_n\coloneqq\{h_{i_1},\ldots,h_{i_d}\}$, that are all 
the possible subsets of $S$ complete and linearly independent. Now, for every $D_n$ we construct 
the dual set $\tilde D_n\coloneqq\{\tilde\vh^{(n)}_1,\ldots,\tilde\vh^{(n)}_d\}$, such that for
$\vh_i\in D_n$ one has $\tilde \vh^{(n)}_j\cdot\vh_{i_k}=\delta_{jk}$. Now, let $\tilde D\coloneqq\bigcup_n\tilde D_n$, and finally we can express the so-called (first) Brillouin zone
as
\begin{align}
B\coloneqq\bigcap_{\vh\in \tilde D}\{\vk\in\Int^d\mid -\pi\leq\tilde \vh\cdot\vk\leq\pi\}
\end{align}
Now, the common (improper) eigenvectors of the right regular representation are
\begin{align}
&\kt\vk\coloneqq\frac1{|B|}\sum_{\vx\in A}e^{-i\vk\cdot\vx}\kt \vx,
\label{eq:planwav}
\end{align}
where we adopted the vector notation $\vx$ for $x\in A$, meaning that if $x=\prod_ih_i^{s_i}$, 
then $\vx=\sum_i s_i\vh_i$, and where $|B|$ is the Borel measure of $B$ in $\Int^d$. It is
easy to verify that
\begin{align}
T_{\vx}\kt\vk=e^{-i\vk\cdot\vx}\kt\vk.
\end{align}
A QW of the form~\eqref{eq:walk} on a Cayley graph of $\Int^d$ can be decomposed in the
representations $\tilde W(\vk)$ supported on $\kt\vk\otimes\Cmp^{|H|}$, as follows
\begin{align}
\tilde W({\vk})=\sum_{\vh_i\in S}e^{-i\vk\cdot\vh_i}W(\vh_i).
\end{align}

One can easily prove that $\tilde W(\vk)$ is unitary for every $\vk\in B$, and this makes it easier
diagonalising $W$, as well as studying its properties. 

\section{Solutions of the unitarity conditions in the three dimensional case}

\begin{figure}
\centering
\includegraphics[width=5cm]{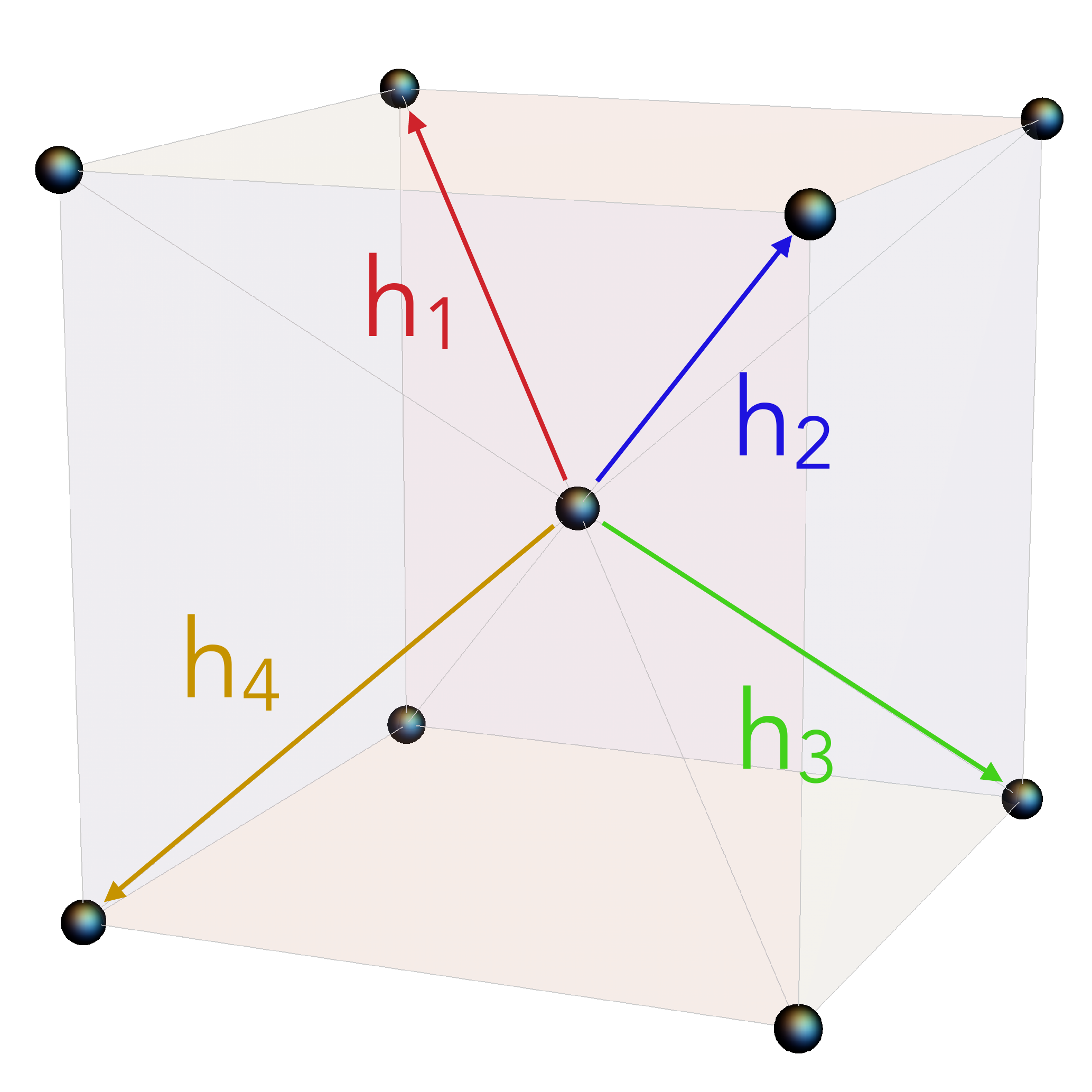}
\caption{The elementary cell of the BCC lattice. The generators $\vh_1$, $\vh_2$, $\vh_3$, $\vh_4$ are higlighted.}
\label{f:bcc}
\end{figure}

The condition of reversibility in the linear Fermionic case requires unitarity of the walk 
operator $W$. In terms of the transition matrices one can express the unitarity conditions as 
follows
\begin{align*}
&\sum_{\vh_i\in S}W^\dag(\vh_i)W(\vh_i)=\sum_{\vh_i\in S}W(\vh_i)W^\dag(\vh_i)=I_{|H|},\\
&\sum_{\vh_i-\vh_j=\vh'}W^\dag(\vh_i)W(\vh_j)=\sum_{\vh_i-\vh_j=\vh''}W(\vh_j)W^\dag(\vh_i)=0
\end{align*}
This system of second degree equations has no easy solution. Exploiting isotropy, one can make 
the above system much easier to solve. In the following, we will consider the three-dimensional
case $d=3$. Isotropy, in the first place, selects a unique possible Cayley graph for a QW on 
$\Int^3$, i.e.~the one represented in $\Reals^3$ by the Body-Centered Cubic (BCC) lattice, with
four generators $\vh_1+\vh_2+\vh_3+\vh_4=0$ (see fig.~\ref{f:bcc}).
Secondarily, one can prove that the isotropic solutions of the above 
equations only four, modulo local unitary equivalence, i.e.~modulo 
changes of basis in the Hilbert space $\Cmp^{|H|}$ of the internal 
degrees of freedom. The solutions are the following, given in terms of 
the matrices $\tilde W(\vk)=W^\pm_\vk$ and $\tilde W(\vk)=Z^\pm_\vk$:
\begin{align}
\begin{aligned}
&W^\pm_\vk=d^\pm_{\vk} I-i\vn_{\vk}^\pm\cdot\vsig^\pm,&&Z_{\vk}^\pm=[W^\pm_\vk]^T,\\
&\vn^\pm_\vk=
\begin{pmatrix}
s_xc_yc_z\mp c_xs_ys_z\\
c_xs_yc_z\pm s_xc_ys_z\\
c_xc_ys_z\mp s_xs_yc_z
\end{pmatrix},
&& d^\pm_\vk=c_xc_yc_z\pm s_xs_ys_z,\\
&s_i\coloneqq\sin k_i,\quad c_i\coloneqq\cos k_i,&& i=x,y,z,
\end{aligned}\label{eq:weylqw}
\end{align}
where $X^T$ denotes the transpose of $X$, $\vsig^+=(\sigma_x,\sigma_y,\sigma_z)$ are the Pauli 
operators, and $\vsig^-=(\sigma_x,-\sigma_y,\sigma_z)$ are their transposes. The above walks are 
called Weyl quantum walks, because in a suitable regime their action can be approximated by the Weyl 
equations.

\begin{figure}[h]
\centering
\includegraphics[width=9cm]{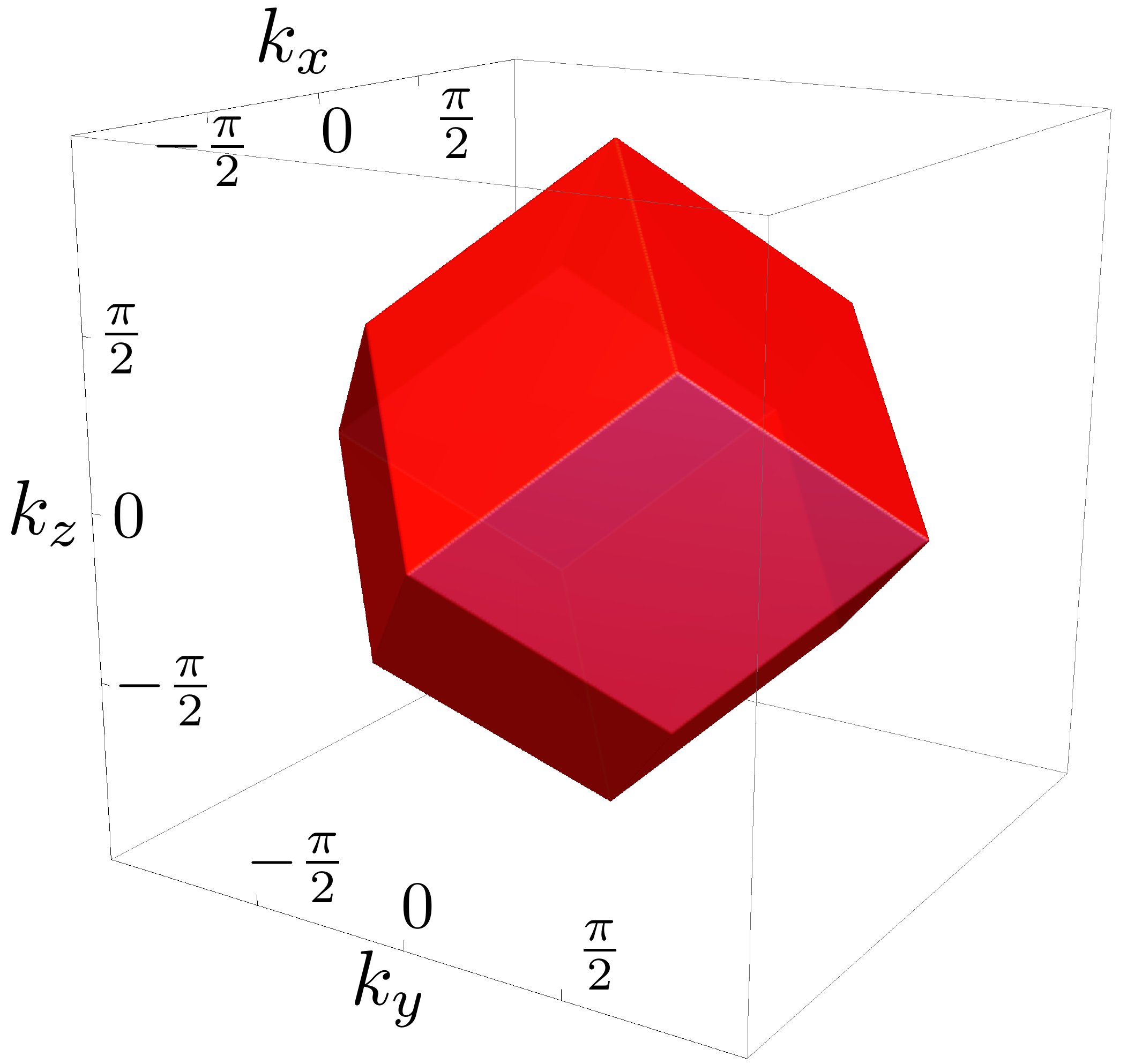}
\caption{The Brillouin zone for the Weyl walks in Eq.~\eqref{eq:weylqw}.}
\label{fig:brill}
\end{figure}

The spectrum of the Weyl walks is now very easily calculated to be 
\begin{align}
\Spec(W^\pm_\vk)=\Spec(Z^\pm_\vk)=\{e^{-i\omega^\pm_\vk},e^{i\omega^\pm_\vk}\},\qquad\omega^\pm_\vk=\arccos d_\vk^\pm.
\end{align}
The Brillouin zone is defined by the following equations
\begin{align}
B=\bigcap_{\substack{i,j\in\{x,y,z\}\\s=\pm1}}\{-\pi\leq k_i+sk_j\leq\pi\}
\end{align}
and it is represented in Fig.~\ref{fig:brill}
The function $\omega^+:B\to[0,\pi)$ defined by $\omega^+:\vk\mapsto\omega^+_\vk$ is the 
{\em dispersion relation} of the QW, and along with the {\em wave vector} $\vk$ and the 
{\em helicity vector} $\vn^\pm_\vk$ contains all the information about its kinematics. 
In particular, a wave-packet with wave-vector distribution peaked around $\vk$ propagates with
a group velocity given by $\vv^\pm_\vk=\vnab\omega^\pm_\vk$.

The transition matrices for the Weyl walks are
\begin{align}
&A_{\vh_1}=
\begin{pmatrix}
\zeta^\pm&0\\
\zeta^\pm&0
\end{pmatrix},
&&A_{-\vh_1}=
\begin{pmatrix}
0&-\zeta^\mp\\
0&\zeta^\mp
\end{pmatrix},\\
&A_{\vh_2}=
\begin{pmatrix}
0&\zeta^\pm\\
0&\zeta^\pm
\end{pmatrix},
&&A_{-\vh_2}=
\begin{pmatrix}
\zeta^\mp&0\\
-\zeta^\mp&0
\end{pmatrix},\\
&A_{\vh_3}=
\begin{pmatrix}
0&-\zeta^\pm\\
0&\zeta^\pm
\end{pmatrix},
&&A_{-\vh_3}=
\begin{pmatrix}
\zeta^\mp&0\\
\zeta^\mp&0
\end{pmatrix},\\
&A_{\vh_4}=
\begin{pmatrix}
\zeta^\pm&0\\
-\zeta^\pm&0
\end{pmatrix},
&&A_{-\vh_4}=
\begin{pmatrix}
0&\zeta^\mp\\
0&\zeta^\mp
\end{pmatrix},
\end{align}
where $\zeta^\pm=\frac{1\pm i}4$. It is now easy to check that the isotropy group is the Heisenberg
group $\Int_2\times\Int_2$ and its projective unitary representation on $\Cmp^2$ is
$\{I,i\sigma_x,i\sigma_y,i\sigma_z\}$.

The corresponding Fermionic CA are called Weyl automata, and we will denote them by the symbol 
$\tW^\pm$.

\section{Small wave-vector regime}

As we mentioned earlier, the name ``Weyl'' quantum walk, or ``Weyl'' automaton is justified by
the behaviour of the solutions of the walk dynamics in a suitable approximation, that we deem 
{\em small wave-vector regime}. This regime is defined by the condition that the state of the
memory array is given by a vector $\kt\psi\in l^2(A)\otimes\Cmp^2$ that is a superposition of plane 
waves $\kt\vk$ with amplitudes narrowly peaked around a value $\vk_0$, i.e.
\begin{align}
\kt\psi=\sum_{j\in\{\pm\}}\int_Bd^3\vk\psi_j(\vk)\kt\vk\kt{u^j_\vk},
\end{align}
where $W_\vk\kt{u^j_\vk}=e^{ij\omega^\pm_\vk}\kt{u^j_\vk}$, and
\begin{align}
|\psi_j(\vk)|<\varepsilon,\quad\forall j,\ \forall \vk:\ |\vk-\vk_0|\geq\delta,
\end{align}
for some $|\vk_0|\ll\pi$. Notice that we did not specify the sign $\pm$ in the expressions $W_\vk$ 
and $\kt{u^j_\vk}$, for the sake of a lighter notation. What we discuss in the following, indeed, 
holds independently of the special QW, unless otherwise specified. The symbol $W_\vk$ can thus 
denote $W^\pm_\vk$ and $Z^\pm_\vk$. In order to analyse the evolution in such regime, let us 
introduce the {\em interpolating Hamiltonian} $H_I(\vk)$, defined by
\begin{align}
W_\vk=\exp\{-i H_I(\vk)\}.
\end{align}
If we embed the discrete lattice $(\vx,t)$ quasi-isometrically in $\Reals^4$, and extend the domain 
of all functions of $\vk$ from $B$ to the whole $\Reals^3$, the operator $H_I(\vk)$ defines a 
Hamiltonian for a continuous-time evolution on 
$L^2(\Reals^3)\otimes\Cmp^2$, given by
\begin{align*}
U(t)=\int_{\Reals^3}d^3\vk\kt\vk\br\vk\otimes\exp\{-i H_I(\vk)t\}.
\end{align*}
The term ``interpolating Hamiltonian'' refers to the fact that, if we apply $U(t)$ to the extension
of a wave packet $\psi(\vk)=(\psi_+(\vk),\psi_-(\vk))$, the solution 
$\psi(\vk,t)=(\psi_+(\vk,t),\psi_-(\vk,t))$ satisfies Schr\"odinger's equation
\begin{align}
i\partial_t\psi(\vk,t)=H_I(\vk)\psi(\vk,t).
\end{align}
Now, let us consider the Weyl automaton $W^\pm_\vk$. For a small wave-vector packet, to first order 
in $|\vk|$ one has $H_I(\vk)=\vk\cdot\vsig$, and the differential equation becomes (modulo a unitary 
transformation in the case of $W^-_\vk$)
\begin{align}
i\partial_t\psi(\vk,t)=\pm\vk\cdot\vsig\psi(\vk,t).
\end{align}
Now, the inverse Fourier transform of the latter are Weyl's equations
\begin{align}
i\partial_t\psi(\vx,t)=\pm\vsig\cdot\vnab\psi(\vx,t).
\end{align}

\section{Reversibility and the Dirac automaton}

Up to now we did only impose the unitarity condition, which is necessary for our strengthened notion
of reversibility, but not sufficient. We observe that actually the Weyl quantum walks are not
equivalent to their inverses through a local transformation acting identically and independently on
every cell. Indeed, such a transformation can only rotate the vector $\vn_\vk^\pm$ by a fixed 
rotation, independent of $\vk$, and this is not sufficient to turn $\vn_\vk^\pm$ into $-\vn_\vk^\pm$
for every $\vk$. However, the construction that we presented in the discussion about reversibility
will produce a reversible automaton that locally behaves as a Weyl automaton: it is sufficient
to double every cell, thus having $|H|=4$, and taking the automaton corresponding to the walk 
operator
\begin{align}
A^\pm\coloneqq W^\pm\otimes (W^\pm)^\dag.
\end{align}
In the small wavevector regime, the above QW behaves as a massless Dirac field.
Notice that one can take one further step, and introduce the family of Dirac automata, defined by
the walk operators
\begin{align}
D^\pm\coloneqq
\begin{pmatrix}
nW^\pm&imI\\
imI&n(W^\pm)^\dag
\end{pmatrix},
\end{align}
where $n^2+m^2=1$. The parameter $m$ parametrises the elements of this family of QWs, each of which
has a small-wavelength behaviour governed by the following differential equation
\begin{align*}
i\partial_t\psi(\vx,t)=(\pm in\valp\cdot\vnab-m\beta)\psi(\vx,t),
\end{align*}
where one can express $\valp=\gamma^0\vgam$, and $b=\gamma^0$ in terms of the Dirac matrices 
$\gamma^\mu$ representing the Clifford algebra for $SU(1,3)$, 
i.e.~$\{\gamma^\mu,\gamma^\nu\}=2\eta^{\mu\nu}$. For $m\ll n$, one can take a further 
approximation, truncating to first order in $m$, and rewrite $\valp$ and $\beta$,
thus obtaining Dirac's equation
\begin{align}
(i\gamma^\mu\partial_\mu-mI)\psi(\vx,t)=0.
\end{align}
We remark that the Dirac QWs correspond to the Dirac CAs which can be obtained as
\begin{align}
\tD^\pm=[\tW^\pm\otimes\tI]\tilde\tS_m[(\tW^{\pm})^{-1}\otimes\tI]\tS,
\end{align}
where $\tilde\tS_m\coloneqq (m\tI+in\tS)\tS$. We also observe that 
\begin{align*}
\tilde\tS_m=\prod_{g\in A}\tilde\tS_m(g),
\end{align*}
where $\tilde\tS_m(g)\coloneqq(m\tI_g+in\tS_g)\tS_g$, and $\tS_g$ denotes the swap operators acting 
on the cell $g$, i.e.~$\tS_g[\psi_s^u(g)]=\psi_s^d(g)$, $\tS_g[\psi_s^d(g)]=\psi_s^u(g)$, $u$ and 
$d$ denoting the ``up'' and ``down'' local modes (the first and second copy of the original memory array, respectively). Let us now define 
\begin{align*}
{\tS^\pm_m}'(g)\coloneqq [\tW^\pm\otimes\tI]\tilde\tS_m[(\tW^{\pm})^{-1}\otimes\tI].
\end{align*}
Since $\tW^\pm\otimes\tI$ is a local automorphism of the Fermionic algebra, one has that i) ${\tS^\pm_m}'(g)$ acts on the neighbourhood of $g$ of the $u$ modes, and on $g$ of the $d$ modes; ii) since $[\tilde\tS_m(g),\tilde\tS_m(g')]=0$, also $[{\tS^\pm_m}'(g),{\tS^\pm_m}'(g')]=0$. 
This implies that $\tD^\pm$ can be decomposed as
\begin{align*}
\tD^\pm=\prod_{g\in A}{\tS^\pm_m}'(g)\prod_{g'\in A}\tS_{g'}.
\end{align*}
Upon defining $N^\pm_g$ the neighbourhood of $g$ in the Cayley graph of the CA, one can then find a 
subset $H\subseteq A$ such that $N^\pm_{h_1}\cap N^\pm_{h_2}=\emptyset$ for $h_1\neq h_2\in H$, and 
such that finitely many translations $Hg_i^{-1}$ of $H$ cover the whole $A$. In this way one can
easily obtain that
\begin{align*}
\tD^\pm=\prod_{i=1}^k\left(\prod_{h\in Hg_i^{-1}}{\tS^\pm_m}'(h)\right)\prod_{g\in A}\tS_g.
\end{align*}
This means that the CAs $\tD^\pm$ admit a decomposition into finitely many (precisely $k$) layers of 
local, non overlapping unitary gates acting locally on the neighbourhoods of a cell, i.e.~they admit 
a so-called {\em Margolus decomposition}~\cite{schumacher2004reversible}.

\section{Theory of light}

We showed that our CAs reproduce in the small wavelength regime Weyl's and Dirac's equations. We now 
discuss how one can reconstruct also Maxwell's equations from the massless Dirac CA, following 
Ref.~\cite{BISIO2016177}. The idea is to study particular entangled states where the 
correlations are finely tuned. The second-quantisation analysis is easier than the QW one, thus, we 
start introducing some notation. First, we denote by $\varphi_s(\vx,t)$ the field operators for the 
``up'' component of the $\vx$ cell at step $t$, and $\psi_s(\vx,t)$ those for the ``down'' one. 
Correspondingly, we will denote by $\varphi_s(\vk,t)$ and $\psi_s(\vk,t)$ the field operators for 
the normal modes $\vk$, i.e.~
$\varphi_s(\vk,t)=1/|B|^{1/2}\sum_{\vx\in A}e^{-i\vk\cdot\vx}\varphi_s(\vx,t)$, 
and $\psi_s(\vk,t)=1/|B|^{1/2}\sum_{\vx\in A}e^{-i\vk\cdot\vx}\psi_s(\vx,t)$. Now, we define the 
following bilinear operators
\begin{align*}
&\vF_T(\vk,t)\coloneqq \vF(\vk,t)-\frac{\vn_{\frac\vk2}}{|\vn_{\frac\vk2}|}\left[\frac{\vn_{\frac\vk2}}{|\vn_{\frac\vk2}|}\cdot\vF(\vk,t)\right]\\
&F^i(\vk,t)\coloneqq \sum_{ss'}\int_{B'} d^3\vq f_\vk(\vq)\varphi_s(\vk/2-\vq,t)\sigma^i_{ss'}\psi_{s'}(\vk/2+\vq,t),\\
&\vE(\vk,t)\coloneqq|\vn_{\frac\vk2}|\{\vF_T(\vk,t)+\vF_T^\dag(\vk,t)\},\\
&\vB(\vk,y)\coloneqq i|\vn_{\frac\vk2}|\{\vF_T(\vk,t)-\vF_T^\dag(\vk,t)\},
\end{align*}
where $B'$ is a suitable domain corresponding to the Brillouin zone rescaled by a factor $1/2$.
One can then verify (see Ref.~\cite{BISIO2016177}) that the operators $\vE$ and $\vB$ 
approximately satisfy Maxwell's equations in the form
\begin{align}
&\partial_t \vE(\vk,t)=i2\vn_{\frac\vk2}\times\vB(\vk,t),\\
&\partial_t \vB(\vk,t)=-i2\vn_{\frac\vk2}\times\vE(\vk,t),\\
&2\vn_{\frac\vk2}\cdot \vE(\vk,t)=0,\\
&2\vn_{\frac\vk2}\cdot \vB(\vk,t)=0.
\end{align}
Moreover, despite the Fermionic nature of the fields that define the operator $\vF(\vk,t)$, one can 
prove that, upon suitable choice of the function $f_\vk(\vq)$, its transverse components 
\begin{align*}
&\gamma^i(\vk)\coloneqq\vu_\vk^i\cdot\vF(\vk,t_0),\quad i=1,2\\
&\vu^i_\vk\cdot\vn_{\frac\vk2}=\vu^1_\vk\cdot\vu^2_\vk=0,\quad|\vu^i_\vk|=1,\quad(\vu^1_\vk\times\vu^2_\vk)\cdot\vn_{\frac\vk2}>0,
\end{align*}
approximately satisfy the canonical commutation relations
\begin{align}
[\gamma^i(\vk),\gamma^j(\vk')]=\delta_{ij}\delta_B(\vk-\vk').
\end{align}

\section{Special relativity and space-time symmetries}\label{sec:symm}

The question that we discuss in the present section regards the emergence of space-time. We already
stressed that the quasi-isometry class of the Cayley graph induced by a CA identifies in a family of
manifolds, where the evolution rule can be described in terms of interpolating dynamics. The 
differential equations that we found, such as Weyl's, Dirac's or Maxwell's, have a common feature:
covariance under the Poincar\'e group. The symmetry group of space-time in this case is not 
previously fixed, but is the result of various approximations. The nature of the emergent space-time
is intrinsically connected with the symmetries of physical laws represented by CAs. However, for 
the
time being, we do not have an operational interpretation for the symmetry group. This can be made 
clearer if we observe that we do not have an acceptable notion of a boost or a rotation 
connecting observers in two different reference frames. Indeed, the Cayley graph is a graphical 
representation of causal connections between events, and the CA represents a physical law. What does 
it mean to ``move'' with respect to a causal connection or with respect to a physical law? The 
detailed treatment of this question can be found in 
Refs.~\cite{refId0,PhysRevA.94.042120,apadula2018symmetries}

To make the role of symmetries consistent we start observing that a CA represents a physical law as
described by an observer. Now, every ``inertial'' observer will describe the evolution of physical 
systems using the same rule. Symmetries are thus required only to map the state of systems as seen 
by one observer to the state of the same systems as seen by a different observer. Now, the
conceivable transformations are once again dictated by our CA. The question is then: how does a CA
enforce a group of transformations that can be interpreted as a change of reference frame?

Here comes into play the relativity principle. IN its original formulation, due to Galileo, the
principle states that {\em every inertial observer describes physical laws in the same way}. Thus,
the maps representing changes of reference frame are those that leave the CA invariant. How does
this simple rule allow us to reconstruct the symmetry group? We proceed in the following way.
First of all, our Fermionic CAs are completely described by QWs. Then, the relativity principle in 
our case will impose that changes of inertial frame must preserve the QW of interest. 

Now, a QW is a unitary operator, and as such it is completely specified by its eigenvalues and 
corresponding eigenvectors. Moreover, an eigenvector of a QW represents a stationary physical 
condition, and every observer has to agree on stationarity. Thus, eigenspaces of the QW must be 
mapped to eigenspaces by a change of inertial frame. Similarly, if two observers change the 
labelling of the Cayley graph in the same way, shifting the ``origin'' labelled by the neutral 
element $e$ to the node formerly labelled $g$, then the change of inertial frame between them 
must be independent of $g$. Let us analyse the consequences of this requirement in detail. If the
origin is shifted from $e$ to $g$, then a node formerly labelled $f$ would now be labelled $gf$.
Thus, the left-regular representation changes accordingly: 
$T_{\vx}\mapsto T_{\vy+\vx}=T_{\vy} T_{\vx}$. This implies that a state that is stationary under 
$T_\vy$ for one of the two observers must be stationary also for the other. This implies that the 
vectors $\kt\vk$ must be mapped to vectors $\kt{\vk'}$. In other words, a change of inertial 
frame must be given by a map $\vk':B\to B$ that maps $\vk\mapsto\vk'(\vk)$, so that the 
eigenvalue equation for the QW $W_\vk$ of interest retains the same solutions
\begin{align}
W_{\vk}\kt{\psi(\vk)}=e^{-i\omega}\kt{\psi(\vk)}.
\end{align}
In order to study the maps $\vk'$ allowed, we first observe that, using unitarity of $W_\vk$, the 
eigenvalue equation can be rewritten as
\begin{align}
(W_{\vk}\pm W_{\vk}^\dag)\kt{\psi(\vk)}=(e^{-i\omega}\pm e^{i\omega})\kt{\psi(\vk)},
\end{align}
namely
\begin{align}
\left\{
\begin{aligned}
&(d_\vk-\cos\omega)\kt{\psi(\vk)}=0,\\
&(\vn_\vk\cdot\vsig-\sin\omega I)\kt{\psi(\vk)}=0.
\end{aligned}
\right.
\end{align}
Since unitarity imposes that $d_\vk=1-|\vn_\vk|^2$, and the second equation requires that 
$\sin\omega=|\vn_\vk|$, the first equation in the system is redundant. Let us then focus on the
second one, and write it as
\begin{align}
&n_\mu(k)\sigma^\mu\kt{\psi(\vk)}=0,\label{eq:disprelcov}\\ 
&n_\mu(k)\coloneqq(\sin\omega,\vn_\vk),\quad\sigma_\mu=(I,\vsig).\nonumber
\end{align}
The form of Eq.~\eqref{eq:disprelcov} clearly recalls a Lorentz-invariant expression. However, 
suppose that a Lorentz transform is applied to the vector $n^\mu(k)$: in general the obtained four-
vector is not in the range of the map $n:B\times[0,2\pi)\to B_1(0)\times[-1,1]\subseteq \Reals^4$.
This problem can be easily overcome multiplying Eq.~\eqref{eq:disprelcov} by a non-vanishing function
$f(k)$ such that the map $k\mapsto f(k)n(k)$ has the same invertibility domains as $k\mapsto n(k)$. 
Defining $p^\mu(k)\coloneqq f(k)n^\mu(k)$, we then have the following expression for the eigenvalue 
equation
\begin{align}
p_\mu(k)\sigma^\mu\kt{\psi(\vk)}=0
\end{align}
Now, given a map $\vk':B\to B$ we can write
\begin{align}
p_\mu(k')\sigma^\mu=e^{ia(k)}\tilde\Gamma_kp_\mu(k)\sigma^\mu\Gamma_k,
\end{align}
for suitable invertible matrices $\tilde\Gamma_k$ and $\Gamma_k$ and a $C^\infty$ map $a:B\times[0,2\pi)\to\Reals$. Clearly the maps $\vk\mapsto\vk'$ 
and $\kt{\psi(\vk)}\mapsto\Gamma_k^{-1}\kt{\psi(\vk)}$ preserve Eq.~\eqref{eq:disprelcov}.
However, this scenario would leave room to an exceedingly wide range of possible changes of inertial 
frame, and thus we impose that the maps $\tilde\Gamma_k$ and $\Gamma_k$ do not actually depend on 
$k$, i.e.~we require
\begin{align}
p_\mu(k')\sigma^\mu=\tilde\Gamma p_\mu(k)\sigma^\mu\Gamma.
\end{align}
One can then prove that the allowed maps $k'$ form then a realisation of the semidirect product of 
the Poincar\'e group and a group of invertible radial diffeomorphisms of the Brillouin zone. 

\begin{figure}[h]
\centering
\includegraphics[width=13cm]{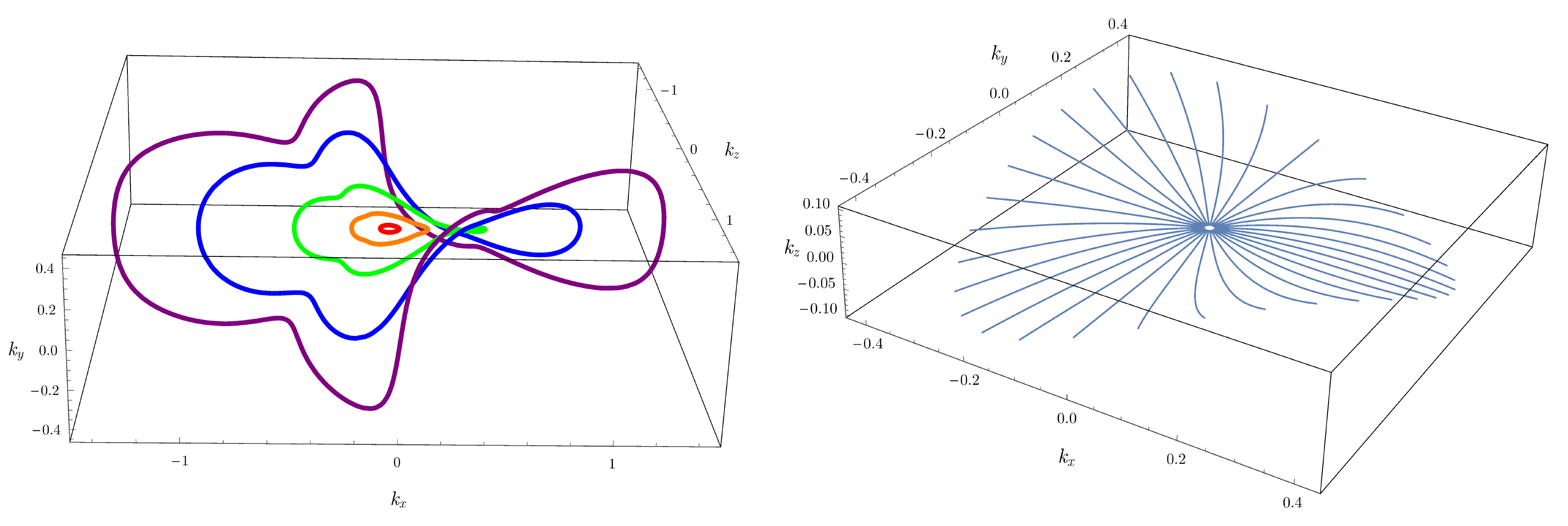}
\includegraphics[width=6cm]{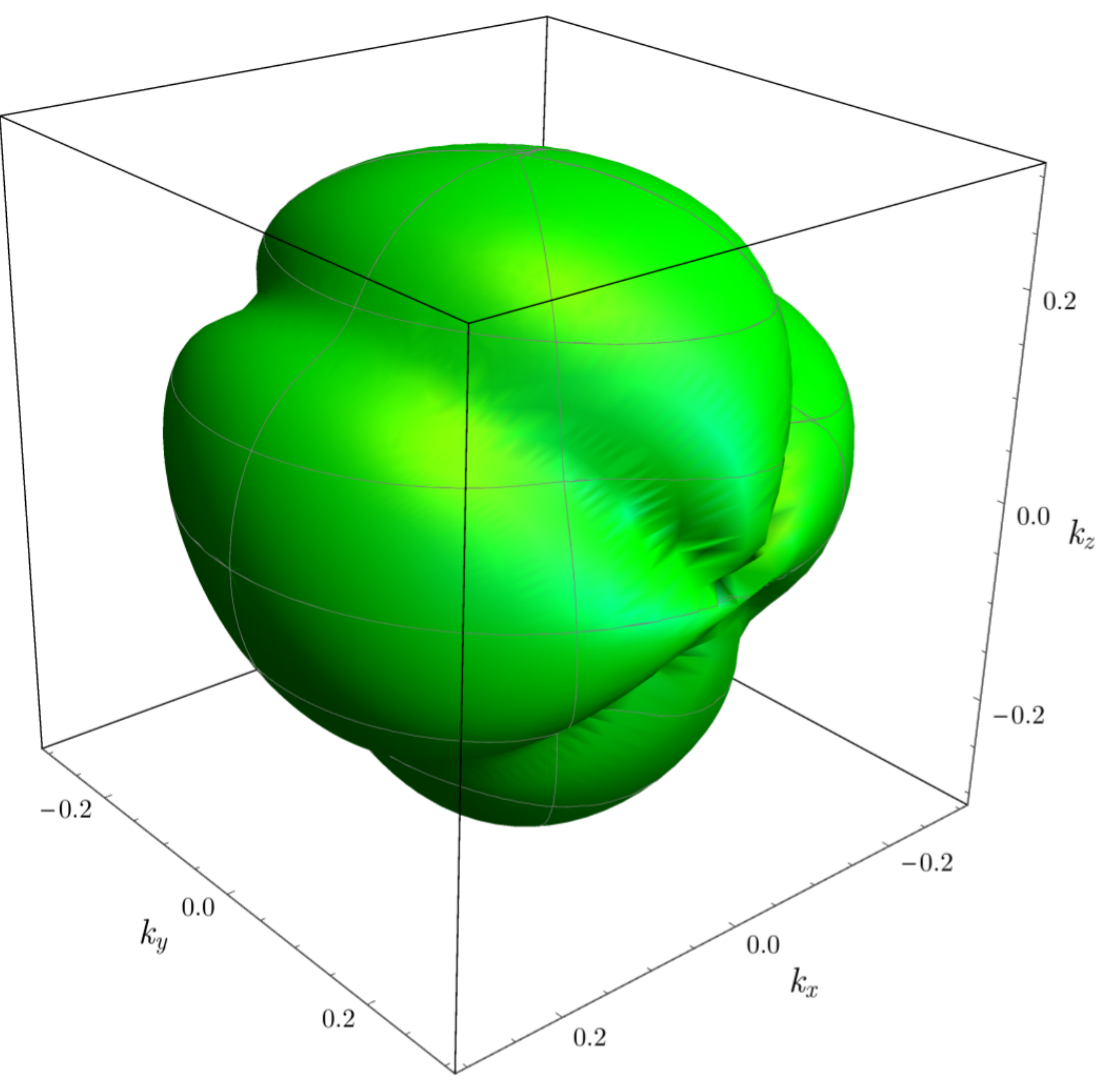}
\caption{Three examples of orbits of a four-vector $k\in B$ under subgroups of the symmetry group of 
the Weyl CA. In the top left figure we show the orbits of various $k$'s under the group of rotations under a fixed axis. Notice the vertical deformation with respect to the usual orbits under rotation. In the top right figure we show the orbits of small $k$'s under boosts in various directions. In the bottom figure, we show the orbit of a vector $k$ under the full group of rotations.}
\label{f:orbits}
\end{figure}

The existence of suitable functions $f$ that allow one to define $p^f_\mu(k)$ is proved in 
Ref.~\cite{PhysRevA.94.042120}. Notice that we made the dependence of the function $p^f_\mu$ on $f$ 
explicit, because in principle there might exist more than one function $f$ (actually, there are 
continuously many). On the other hand, one can find the detailed derivation of the full group in 
Ref.~\cite{Bisio:2017aa}. In Fig.~\ref{f:orbits} we show some orbits of vectors $k$ under the subgroups of rotations, rotations around a fixed axis, and boosts.

In Ref.~\cite{apadula2018symmetries} the symmetries of the massive (Dirac) QWs in $1+1$-d are studied 
in the same approach, and it is shown that in that case the changes of inertial reference frame form 
a group that is a semidirect product as above, but involving the linear group $SO(1,2)$, 
i.e.~recombining the mass parameter with the components of $k$.

All the realisations of the aforementioned symmetry groups provide non-linear diffeomorphisms of the
Brillouin zone. As a consequence, there is no dual transformation on space-time coordintes, but 
rather a different transformation for every value of $k$. This phenomenon is known in the field of 
doubly special relativity, an approach to quantum gravity based on the relaxation of the usual 
Lorentz symmetry that allows one a space-time symmetry group that preserves, besides the speed ofg 
light $c$, also a length (or equivalently an energy), that could be the Planck length $l_P$. Its most 
counterintuitive consequence of this kind of deformations leads to the phenomenon deemed {\em 
relative locality}: the space-time transformations describing the same change of inertial frame are 
different for systems in different regimes of the energy spectrum. As a result, events that coincide 
in one reference frame might not coincide in another one. Such a possibility was studied for the Weyl 
QW in 1+1-d in Ref.~\cite{refId0}.

\section{Conclusion}

The success of the program that we discussed so far is very promising. We recovered the dynamics of 
free relativistic quantum fields and the Poincar\'e symmetry of space-time. To this end, a remark is
in order: the dimension 3+1 is not derived, but assumed from the beginning. The future developments 
of the present approach require the construction of interacting theories from suitable principles,
and an approximation toolkit in order to perform calculations for the sake of comparison with the 
standard model, along with prediction of possible deviations. In this respect, a quantum version of 
the gauge principle seems very appropriate in the context of our approach where time is treated as
a discrete variable. In this case, indeed, there is no way to compare the ``canonical basis'' of 
the local Hilbert space at any site at step $t$ with that at step $t+1$, and this implies that 
freedom must be allowed in the definition of such basis at every step, in a way that must not break
homogeneity, nor create ``particles'' out of nothing. This requirement leads in 1+1-d to an 
essentially unique family of non-linear interactions that were studied in 
Ref.~\cite{PhysRevA.97.032132}, where the model was analytically diagonalised in the two-fermion 
sector.

The study of symmetries in the presence of interactions is of crucial importance in view of the 
challenge of recovering a space-time geometry that depends on the state of the fields, in the 
perspective of reconciling a dynamic geometry with the quantum nature of elementary systems. The 
latter problem clearly exhibits very hard challenges, and tackling simplified versions of it, like 
those reported in Section~\ref{sec:symm}, allows us to fill the gaps in our present understanding.

\bibliographystyle{unsrt}
\bibliography{ilasl}

\end{document}